\documentclass{article}

 \usepackage[preprint]{neurips_2026}

\usepackage[utf8]{inputenc} 
\usepackage[T1]{fontenc}    
\usepackage{hyperref}       
\usepackage{url}            
\usepackage{booktabs}       
\usepackage{amsfonts}       
\usepackage{nicefrac}       
\usepackage{microtype}      
\usepackage[table]{xcolor}

\usepackage{amsmath}
\usepackage{amssymb}
\usepackage{mathtools}
\usepackage{amsthm}
\usepackage{wrapfig}
\usepackage{algorithm}
\usepackage{algorithmic}

\usepackage{multirow}
\usepackage{xspace}
\usepackage{makecell}
\usepackage{enumitem}

\usepackage{amsmath,amsfonts,bm}

\def\eqref#1{equation~\ref{#1}}

\def\1{\bm{1}}

\def\rva{{\mathbf{a}}}

\def\rve{{\mathbf{e}}}

\def\rvh{{\mathbf{h}}}
\def\rvu{{\mathbf{i}}}

\def\rvs{{\mathbf{s}}}

\def\rvu{{\mathbf{u}}}
\def\rvv{{\mathbf{v}}}

\def\rvx{{\mathbf{x}}}

\def\rvz{{\mathbf{z}}}

\def\rmB{{\mathbf{B}}}

\def\rmP{{\mathbf{P}}}

\def\rmX{{\mathbf{X}}}

\DeclareMathAlphabet{\mathsfit}{\encodingdefault}{\sfdefault}{m}{sl}
\SetMathAlphabet{\mathsfit}{bold}{\encodingdefault}{\sfdefault}{bx}{n}

\def\gD{{\mathcal{D}}}
\def\gE{{\mathcal{E}}}

\def\gG{{\mathcal{G}}}

\def\gI{{\mathcal{I}}}

\def\gT{{\mathcal{T}}}

\newcommand{\E}{\mathbb{E}}

\newcommand{\R}{\mathbb{R}}

\newcommand{\method}{ATMOS\xspace}

\title{Atomic Trajectory Modeling with State Space Models for Biomolecular Dynamics}

\author{%
  Liang Shi\thanks{Work conducted during their internships at BioGeometry.} \\
  School of Computer Science \\
  Peking University \\
  \texttt{liang.shi@stu.pku.edu.cn} \\
  \And
  Jiarui Lu \\
  Mila - Qu\'ebec AI Institute \\
  Universit\'e de Montr\'eal \\
  \texttt{jiarui.lu@umontreal.ca} \\
  \And
  Junqi Liu\footnotemark[1] \\
  School of Computer Science \\
  Peking University \\
  \texttt{liujunqi@pku.edu.cn} \\
  \AND
  Chence Shi \\
  BioGeometry \\
  \texttt{chence.shi@umontreal.ca} \\
  \And
  Zhi Yang\thanks{Corresponding authors.} \\
  School of Computer Science \\
  Peking University \\
  \texttt{yangzhi@pku.edu.cn} \\
  \And
  Jian Tang\footnotemark[2] \\
  Mila - Qu\'ebec AI Institute \\
  HEC Montr\'eal \\
  CIFAR AI Chair \\
  \texttt{jian.tang@hec.ca} \\
}

\begin{document}

\maketitle

\begin{abstract}
Understanding the dynamic behavior of biomolecules is fundamental to elucidating biological function and facilitating drug discovery.  
While Molecular Dynamics (MD) simulations provide a rigorous physical basis for studying these dynamics, they remain computationally expensive for long timescales.  
Recent deep generative models accelerate conformation generation but often either discard temporal correlations entirely or struggle to condition on the extended history required for faithful kinetics, a consequence of the effectively non-Markovian nature of partially observed subsystem coordinates.  
To bridge this gap, we introduce \method, a novel generative framework based on State Space Models (SSM) designed to generate atom-level MD trajectories for biomolecular systems. 
\method integrates a Pairformer-based state transition mechanism to capture temporal dependencies, with a diffusion-based module to decode trajectory frames autoregressively.
We demonstrate that \method achieves state-of-the-art performance in generating conformation trajectories for both protein monomers and protein-ligand systems.  
This work provides a unified and computationally efficient framework for biomolecular trajectory generation, taking a step toward dynamics foundation models.
\end{abstract}

\section{Introduction}\label{intro}
Elucidating the functional mechanisms of biomolecules requires a comprehensive understanding of protein dynamics, rather than merely analyzing static crystal structures. Traditionally, Molecular Dynamics (MD) simulations are the gold standard for this task, relying on fundamental physical principles to reveal critical thermodynamic and kinetic properties. However, despite rigorous adherence to physical laws, traditional MD methods are computationally expensive, often struggling to access biologically relevant timescales or sample rare conformational transitions efficiently.

To alleviate these computational burdens, deep generative models have emerged as powerful surrogates. Broadly, these approaches fall into two categories. Static ensemble generators, such as AlphaFlow~\cite{jing2024alphaflow} and BioEmu~\cite{lewis2025bioemu}, learn to approximately sample conformations from the equilibrium Boltzmann distribution, yet they do so in an \textit{i.i.d.} fashion, discarding temporal correlations and kinetic information. In contrast, trajectory generation methods explicitly model temporal evolution: MDGen~\cite{jing2024mdgen} applies flow matching to time-series of frames in MD simulations; TEMPO~\cite{xu2025tempo} decouples slow and fast motions hierarchically to sample trajectories; ConfRover~\cite{shen2025confrover} employs a causal transformer with an SE(3) diffusion decoder; and STAR-MD~\cite{shoghi2026scalable} leverages joint spatio-temporal attention for more scalable autoregressive modeling. Despite these advances, existing methods still face challenges. Methods that employ temporal attention scale quadratically with trajectory length. More efficient non-attention approaches~\cite{schreiner2023ito, klein2023timewarp, costa2024equijump} face a different challenge: MD trajectories of a subsystem (e.g., protein-only coordinates) are effectively non-Markovian due to implicit solvent coupling and coarse-graining~\cite{shen2025confrover}. Consequently, accurate kinetic modeling requires conditioning on an extended prefix of past frames, which these methods often struggle to incorporate efficiently.

In this work, we introduce \textbf{A}tomic \textbf{T}rajectory \textbf{MO}deling with \textbf{S}SMs~(\textbf{\method}), which formulates biomolecular dynamics generation as a sequence modeling problem. \method employs a state space model (SSM) as a learnable propagator for MD trajectories. The SSM maintains a recurrent latent state that accumulates information from past frames, enabling the model to condition on an extended temporal context without the quadratic cost of attention. This design directly addresses the non-Markovian nature of the observed subsystem coordinates, as the latent state can efficiently summarize the history required for accurate kinetic predictions. We parameterize the SSM transition function with a Pairformer module~\cite{abramson2024af3}, and decode the updated latent state back to full atomic coordinates via a diffusion model. A hierarchical coarse-to-fine sampling strategy, augmented with chunk stitching, ensures global consistency and temporal smoothness across long roll-outs. Being trained end‑to‑end on MD trajectories, \method achieves state-of-the-art trajectory generation performance on both mdCATH~\cite{mirarchi2024mdcath} and MISATO~\cite{siebenmorgen2024misato} benchmarks. \method provides a unified and computationally efficient framework for biomolecular trajectory generation, taking a step toward protein dynamics foundation models.

\textbf{Main contributions.}
(i) We propose \method, a unified generative framework for biomolecular dynamics that handles diverse systems, ranging from monomeric proteins to protein-ligand complexes.
(ii) We innovatively adapt State Space Models to the trajectory generation, enabling the capture of temporal dependencies with improved inference efficiency compared to traditional simulation or purely attention-based approaches. A hierarchical coarse-to-fine sampling strategy with chunk stitching further ensures long-range temporal smoothness.
(iii) \method achieves state-of-the-art performance on the large-scale mdCATH~\cite{mirarchi2024mdcath} and MISATO~\cite{siebenmorgen2024misato} benchmarks for trajectory generation, validating the model's ability to generalize across diverse biological systems. 

\section{Related Work}

\paragraph{Generative Modeling of Protein Ensembles.}
Deep generative models have been developed to sample conformations \textit{i.i.d.}. Seminal work such as Boltzmann Generators~\cite{noe2019bg} utilized normalizing flows to sample from the Boltzmann distribution, though scaling to systems beyond training remained challenging. Recent advances leverage diffusion and flow matching on large-scale MD datasets; for instance, AlphaFlow~\cite{jing2024alphaflow} fine-tunes pre-trained folding models to capture ensemble diversity, while EBA~\cite{lu2025eba} enhances this framework with physical energy-based alignment. Other approaches focus on specific physical constraints or data regimes: Distributional Graphormer (DiG)~\cite{zheng2024dig} and ConfDiff~\cite{wang2024confdiff} incorporate energy or force guidance to improve physical validity, Str2Str~\cite{lu2024strstr} employs a zero-shot translation framework without reliance on MD training data, ESMDiff~\cite{lu2025slm} adopts latent language modeling to efficiently sample conformation ensembles, and BioEmu~\cite{lewis2025bioemu} integrates experimental stability data to predict thermodynamic properties.

\paragraph{Generative Modeling of Molecular Trajectories.}
While ensemble models capture time-independent distributions, trajectory generation methods aim to emulate the temporal evolution of molecular systems, serving as surrogates for expensive Molecular Dynamics (MD) simulations. Initial efforts, such as ITO~\cite{schreiner2023ito}, TimeWarp~\cite{klein2023timewarp} and EquiJump~\cite{costa2024equijump}, focused on learning time-coarsened transition operators to jump over large time steps. More recent frameworks model the joint distribution of entire trajectories to capture long-range temporal dependencies. MDGen~\cite{jing2024mdgen} treats trajectories as time-series of 3D structures, utilizing flow matching to enable forward simulation, transition path interpolation, and upsampling. To handle the multi-scale nature of protein motions, TEMPO~\cite{xu2025tempo} introduces a hierarchical trajectory generation framework that separates slow collective motions from fast local fluctuations. ConfRover~\cite{shen2025confrover} employs a causal transformer with an SE(3) diffusion decoder to simultaneously learn conformational distributions and dynamics. STAR-MD~\citep{shoghi2026scalable} introduces an efficient joint spatio-temporal attention mechanism, enabling more scalable autoregressive dynamics modeling. Other specialized architectures include BioMD~\cite{feng2025biomd} for protein-ligand unbinding pathways, and PTraj-Diff~\cite{xu2025ptraj-diff} for protein-protein complex dynamics.

\section{Method}\label{method}

\begin{figure*}[t]
    \centering
    \includegraphics[width=1.0\linewidth]{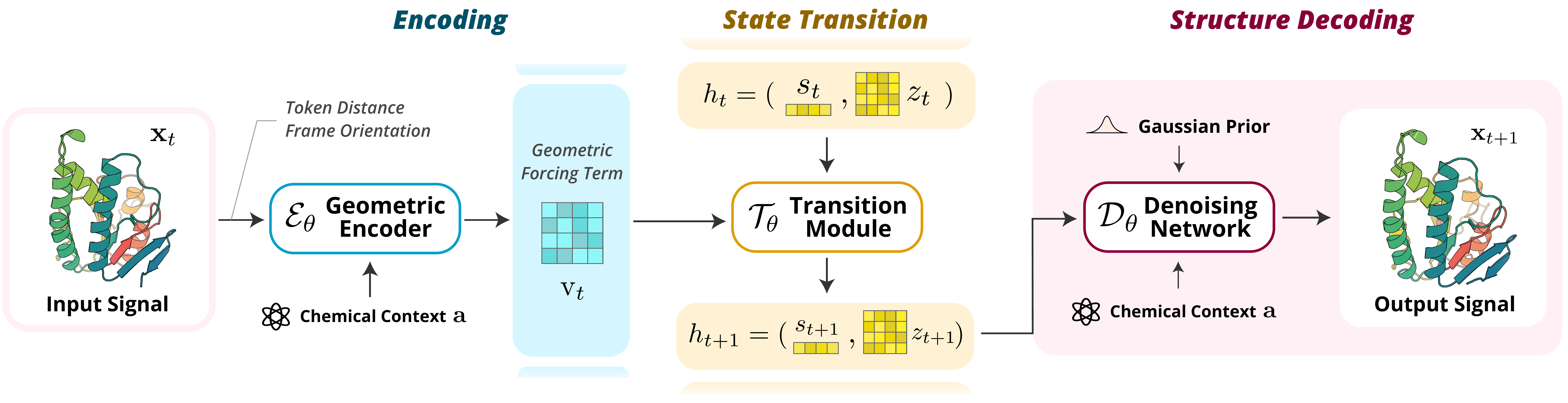}
    \vspace{-2mm}
    \caption{\textbf{Overview of the Proposed Atomic Trajectory Modeling Framework.}}
    \label{fig:main}
\end{figure*}

\subsection{Preliminaries}

\textbf{Problem Formulation.} 
We conceptualize the generation of biomolecular trajectory as a conditional sequence modeling task. Consider a biomolecular system consisting of $N$ atoms. We represent the dynamic evolution of this system as a trajectory of coordinates $\rmX = (\rvx_1, \rvx_2, \dots, \rvx_T) \in \mathbb{R}^{T \times N \times 3}$, where $\rvx_t \in \mathbb{R}^{N \times 3}$ denotes the 3D Cartesian coordinates of all $N$ atoms at time step $t$, and $T$ represents the trajectory length, i.e., the number of frames. 
In addition to coordinates, the system is defined by time-invariant atomic features $\rva$, where $\rva \in \mathbb{R}^{N \times d_a}$ represents atom-level attributes including atom type, residue type, molecule-level identifiers that indicate whether an atom belongs to a protein or ligand, etc. Our objective is to learn a generative model $p_\theta$ that approximates the true distribution of trajectories given the initial conformation $\rvx_1$ and the context $\rva$. Following standard autoregressive practices, we factorize the joint probability over the temporal dimension:
\begin{equation}
    p_\theta(\rmX | \rvx_1, \rva) = \prod_{t=1}^{T-1} p_\theta(\rvx_{t+1} | \rvx_{1:t}, \rva)
\end{equation}
This formulation allows the model to generate physically consistent dynamics by conditioning the next state on the history of the trajectory and the chemical identity of the system.

\textbf{State Space Models (SSM).} 
State Space Models~\cite{gu2021efficiently, gu2021combining} provide a unified framework for modeling sequential data by mapping a 1-dimensional input signal $u(t) \in \mathbb{R}$ to an output signal $y(t) \in \mathbb{R}$ through a latent state $h(t) \in \mathbb{R}^D$. In their continuous-time form, linear SSMs are defined by the following ordinary differential equations (ODEs):
\begin{equation} \label{eq:ssm}
    \dot{h}(t) = \mathbf{A}h(t) + \mathbf{B}u(t), \quad y(t) = \mathbf{C}h(t)
\end{equation}
where $\mathbf{A} \in \mathbb{R}^{D \times D}$, $\mathbf{B} \in \mathbb{R}^{D \times 1}$, and $\mathbf{C} \in \mathbb{R}^{1 \times D}$ are learnable projection matrices. To apply this framework to discrete sequence data (such as MD trajectories sampled at fixed intervals $\Delta$), the continuous system is discretized, typically via the Zero-Order Hold method, yielding the recurrence:
\begin{equation} \label{eq:ssm_zoh}
    h_t = \overline{\mathbf{A}}h_{t-1} + \overline{\mathbf{B}}u_t, \quad y_t = \mathbf{C}h_t
\end{equation}
Here, $\overline{\mathbf{A}}$ and $\overline{\mathbf{B}}$ are the discretized state parameters. This recurrent formulation allows for efficient $O(1)$ inference per step, making SSM particularly well-suited for generating long molecular trajectories compared to the quadratic complexity of standard attention mechanisms.

\textbf{Notations.} 
Throughout this paper, we use bold lowercase letters (e.g., $\rvx, \mathbf{h}$) for vectors and atom-level representations, and bold or calligraphic uppercase letters (e.g., $\mathbf{A}, \Phi, \gE$) for matrices and operations. We denote the coordinate of the $i$-th atom at time $t$ as $\rvx_{t,i} \in \mathbb{R}^3$. The set of all atoms is indexed by $i \in \{1, \dots, N\}$, and discrete time steps are indexed by $t \in \{1, \dots, T\}$.

\subsection{An Atomic Trajectory Modeling Framework}

\textbf{SSM as Learnable Biomolecular Simulators.}
We frame the generation of biomolecular trajectories as the learning of a time-continuous evolution function, functionally analogous to the integrator used in classical MD simulations. In traditional MD simulations, the system evolves according to an integrator that explicitly calculates future positions based on current coordinates, momenta, and inter-atomic forces derived from a force field. While rigorous, this process is computationally intensive for long timescales. To overcome this, \method adopts SSM as efficient, learnable biomolecular propagators. Unlike standard attention-based sequence models that must revisit the entire history for every prediction, SSM maintains a latent recurrent state. This aligns with the physical reality of molecular dynamics: the future evolution of a system is governed by its current \textit{phase-space} state (positions and momenta). By replacing the expensive numerical integrator with a neural SSM, we achieve a framework that is both physically intuitive and computationally efficient.

\textbf{Specifications.}
Formally, we instantiate the SSM framework for atomic trajectory modeling by defining the input signal, latent state, and output prediction as follows:
\begin{itemize}[leftmargin=0.5cm]
    \item \textbf{Input Signal ($\rvx_t$):} The time-varying input is the configuration of the system at time $t$, represented by the atomic coordinates $\rvx_t \in \mathbb{R}^{N \times 3}$. This serves as the observable states of the sequence.
    \item \textbf{Latent State ($\rvh_t$):} The recurrent hidden state $\rvh_t$ acts as an implicit representation of the system's complete phase-space state. While the input $\rvx_t$ provides only the geometric configuration, the latent state $\rvh_t$ captures the auxiliary dynamic variables necessary for propagation. 
    \item \textbf{Output Signal ($\rvx_{t+1}$):} Conditioning on the updated latent $\rvh_{t+1}$, the model operates autoregressively to predict the updated coordinates $\rvx_{t+1} \in \mathbb{R}^{N \times 3}$ for the subsequent time step.
\end{itemize}

\textbf{Learning Objective.}
This framework provides two critical advantages. First, the constant inference complexity of SSMs per step (in recurrent mode) allows \method to generate trajectories of arbitrary length $T$ with linear scaling, addressing the bottleneck of existing autoregressive methods. Second, the continuous-time formulation of SSMs naturally handles the temporal correlations inherent in physical motion, smoothing the modeling of conformational transitions.
Under this framework, \method can be trained to maximize the likelihood of the ground-truth trajectory $\rmX$ given the chemical context $\rva$. The learning objective is:
\begin{equation}
    \mathcal{L}_{\text{SSM}} = - \sum_{t=1}^{T-1} \log p_\theta(\rvx_{t+1} \mid \mathbf{h}_t, \rvx_t, \rva)
\end{equation}
where $p_\theta$ is parameterized by the predicted shift in coordinates.

\subsection{Model Architecture}\label{method:arch}

\textbf{Compact Coupled Latent Representation.}
We adopt a compact \textit{coupled latent} widely used in modern structure prediction architectures~\cite{jumper2021highly, abramson2024af3}.
Specifically, we define the latent SSM state $\rvh_t$ as a coupled representation operating on a tokenized sequence of length $L$ (where $L \le N$): 
\begin{equation}
    \rvh_t \triangleq (\rvs_t, \rvz_t).
\end{equation}
Here, the system is tokenized such that atom-level representations of protein are grouped into residue-level latents, otherwise the atomic resolution is retained. Consequently, $\rvs_t \in \mathbb{R}^{L \times d_s}$ denotes the \textit{single representation}, capturing the state of each residue or ligand atom, and $\rvz_t \in \mathbb{R}^{L \times L \times d_z}$ denotes the \textit{pair representation}, encoding the interactions between tokens. 

\textbf{Neural SSM Parametrization.}
While classical SSMs rely on linear projections (Eq.~\ref{eq:ssm}), modeling complex molecular interactions requires high-capacity non-linear transitions. Based on the coupled latent space, one can effectively parameterize the state evolution using neural networks. The architecture is composed of three functional modules:

\textit{Context and Input Encoding.}
We first initialize the latent state by tokenizing the chemical context $\rva$. An embedding layer maps contexts such as atom types and residue indices to the initial state $\rvh_0 \triangleq(\rvs_0, \rvz_0)=\rve_\theta(\rva)$, establishing the context prior of the system before dynamics begin.
At each time step $t$, the input signal $\rvx_t$ (current atomic coordinates) must be injected into the latent space to update the system's trajectory. We employ a geometric encoder $\gE_{\theta}$ to extract rot-trans invariant features from $\rvx_t$:
\begin{equation}
    \rvv_t = \gE_{\theta}(\rvx_t)
\end{equation}
This term $\rvv_t$ acts as the forcing term (analogous to $\rmB \rvu_t$ in Eq.~\ref{eq:ssm_zoh}), representing the instantaneous geometric update derived from the current frame $\rvx_t$.


\textit{State Transition.}
The core propagator of \method is the state transition function $\gT_\theta$, which evolves the latent state $\rvh$ from $t$ to $t+1$. We instantiate this transition using the Pairformer module~\cite{abramson2024af3}, denoted as $\Phi_\theta: \rvs,\rvz \mapsto \rvs', \rvz' \in \R^{L \times d_s}, \R^{L \times L \times d_z}$. 
The discretized update rule is defined as ($\Delta t >0$ indicates the time step):
\begin{equation}
    \rvh_{t+1} \triangleq  (\rvs_{t+1}, \rvz_{t+1}) = \gT_{\theta}(\rvh_t, \rvv_t, \Delta t).
\end{equation}
Here, the transition kernel $\gT_\theta$ is defined as 
\begin{equation}
    \gT_{\theta}\bigl(\rvh_t, \rvv_t, \Delta t) \triangleq \gT_{\theta}\bigl(\rvs_t, \rvz_t; \rvv_t; \Delta t) = \notag \Phi_\theta\bigl(\rvs_t + \tau_\theta(\Delta t),\;  \rvz_t + \rvv_t + \tau_\theta(\Delta t) \bigr),
\end{equation}
where $\tau_\theta(\cdot)$ is broadcasted learnable timestep embeddings.
Intuitively, the Pairformer $\Phi_{\theta}$ functions as a learnable integrator, calculating the next latent state by reasoning over the previous memory $\rvh_t$ and the current geometric update $\rvv_t$.

\textit{Structure Decoding.}
Finally, to map the evolved latent state back to the coordinate space, we employ a diffusion-based decoder. We model the generation of the next frame $\rvx_{t+1}$ as a reverse diffusion process. The core component is a denoising network $\gD_{\theta}$ that predicts the clean, denoised structure from a noisy state. Specifically, for a given trajectory step $t+1$, we introduce a diffusion time variable $\gamma \in [0, +\infty]$ to distinguish the diffusion time from the trajectory index $t$. The decoder is conditioned on the noisy structure $\tilde{\rvx}^{(\gamma)}$, the chemical context $\rva$, and the evolved latent representations $(\rvh_{t+1})$. The denoising operation is defined as:
\begin{equation}
    \hat{\rvx}_{t+1} = \gD_{\theta}(\tilde{\rvx}^{(\gamma)}, \rva, \rvh_{t+1}, \gamma)
\end{equation}
where $\hat{\rvx}_{t+1}$ represents the predicted denoised coordinates. 

\subsection{Training Objectives}\label{method:training_objectives}

\textbf{Training Regime.}
To efficiently train \method on MD trajectory data, we employ a teacher-forcing~\cite{williams1989learning} with subsampling strategy during the training stage. Rather than unrolling the full trajectory, we optimize the model on sub-trajectory windows and input the prefix states as the ground truth.
Specifically, for a given training trajectory $\rmX = (\rvx_1, \dots, \rvx_T)$ of length $T$, we sample a start index $t_0$, a temporal stride $\Delta t$, and a window size $k$ such that $1 \le t_0 < t_0 + k\Delta t \le T$. The model processes the first $k-1$ frames as context to predict the target state $\rvx_{t_0+k\Delta t}$ at the strided $k$-th step. 
For simplicity, we adopt a fixed stride in this work, whereas we believe it is worth exploring learn dynamics across varied timescales by randomizing $\Delta t$.

\textbf{Loss Functions.} We train \method end-to-end with two important loss terms: (1) latent fidelity loss (distogram) and (2) reconstruction loss (diffusion), which are detailed as follows.

\textit{Latent Fidelity Loss (Distogram).}
To enforce geometric consistency in the latent space, we impose a distogram-based latent fidelity loss. Following AlphaFold 3~\cite{abramson2024af3}, a lightweight prediction head $\rmP_{\theta}$ is applied to the pair representation $\rvz_{t}$, which outputs a 64‑bin probability distribution over discretized distances for each token pair $(i,j)$. The distogram loss is defined as the cross‑entropy between the predicted distribution and the target bin (one-hot) derived from the target coordinates:
\begin{equation}
    \mathcal{L}_{\text{disto}} = - \E_{\rvx_{t}} \big[\frac{1}{L^2} \sum_{i,j=1}^{L} \sum_{b=1}^{64}
    \mathbf{\delta}_{(i,j)}^{(b)}(\rvx_{t}) \,
    \log \rmP_{\theta, (i,j)}^{(b)}(\rvz_{t, (i,j)})\big],
\end{equation}
where $\mathbf{\delta}_{(i,j)}^{(b)}(\rvx_{t})$ indicates whether the distance between tokens $i$ and $j$ in the target structure falls into the bin $b$, and the target frame $\rvx_t$ is sampled via the training regime above.

\textit{Reconstruction Loss (Denoising Diffusion).}
This objective optimizes the diffusion-based decoder $\gD_\theta$ to (conditionally) recover the target frame $\rvx_{t}$ from a noise-corrupted state $\tilde{\rvx}^{(\gamma)}$. Firstly, we parameterize the decoder $\gD_{\theta}$ using an EDM-style diffusion process~\cite{karras2022elucidating}. During training, the noisy state is sampled following the diffusion time schedule $\gamma \sim p(\gamma) \in \R^+$:
\begin{equation}
    \tilde{\rvx}^{(\gamma)} = \rvx_{t} + \gamma \cdot \mathbf{\epsilon}, \quad \epsilon \sim \mathcal{N}(0, \mathbf{I}).
\end{equation}
To ensure the model focuses on non-trivial structural deviations rather than global rotations or translations, we apply the Kabsch algorithm to rigid align the target frame to the denoise coordinates. The loss is defined as:
\begin{equation}
    \mathcal{L}_{\text{recon}} = \mathbb{E}_{\rvx_t, \gamma, \epsilon} \left[ \sum_{i=1}^{N} \omega(i, \sigma_{\text{data}}, \gamma) \left\| \hat{\rvx}_{i} - \text{Align}(\rvx_{t, i}) \right\|^2 \right]
\end{equation}
where $\hat{\rvx} = \gD_\theta(\tilde{\rvx}^{(\gamma)}, \rva, \rvh_t, \gamma)$ is the predicted denoised structure. The weighting term $\omega(i, \sigma_{\text{data}}, \gamma) = w_{\text{entity}}(i) \cdot  w_{\text{diff}}(\gamma, \sigma_{\text{data}})$ balances the loss by accounting for: atom entities $w_{\text{entity}}$ (distinguishes atoms from protein, ligand, etc.),  dataset-specific noise scales $\sigma_{\text{data}} > 0$, and diffusion weighting $w_{\text{diff}}(\gamma, \sigma_{\text{data}})\triangleq (\gamma^2 + \sigma_{\text{data}}^2)/ (\gamma \cdot \sigma_{\text{data}})^2$.

The final training objective is a linear combination as:
\begin{equation}
    \mathcal{L} = \lambda_{\text{recon}} \cdot \mathcal{L}_{\text{recon}} \;+\; \lambda_{\text{disto}} \cdot \mathcal{L}_{\text{disto}},
\end{equation}
with $\lambda_{\text{recon}}>0, \lambda_{\text{disto}}>0$ as tunable hyperparameters.

\subsection{Trajectory Sampling} \label{subsec:trajectory_sampling}

\textbf{Bi-Level Sampling Protocol.}
During inference, \method samples trajectories in a specific temporal granularity $\Delta t$. However, autoregressive generation of high-dimension conformation over long horizons can lose consistency due to accumulated drift. Motivated by prior work~\cite{henzler2007dynamic, xu2025tempo}, we adopt a hierarchical, coarse-to-fine sampling strategy that decouples conformation changes from local fluctuations.
In practice, we specialize two variants of \method: a \textit{keyframe generator} $\gG$ trained with a large temporal stride ($\Delta t^{\text{kf}} \triangleq u\cdot \Delta t$, where $u>1$ is an upsampling factor) to predict distinct anchor conformations, and an \textit{interpolator} $\gI$ trained with a fine temporal stride ($\Delta t$) to fill in the dynamics between anchors; in other words, the interpolator upsamples (with the factor $u>1$) the trajectory generated from $\gG$. 


\textbf{Conditioned Interpolation with Chunk Stitching.}
While the keyframe generator operates via standard autoregressive rollout, the interpolator must produce smooth transitions both between and across anchor intervals. To achieve this, the interpolator variant $\gI$ is conditioned not only on the starting anchor frame $\rvx_t$ and the destination anchor frame $\rvx_{t'}$ (with $t' = t + \Delta t^{\text{kf}}$), but also on a short prefix of fine‑grained frames from the preceding interpolation interval. Concretely, let $\rvx_{t-p\Delta t : t}$ denote the $p$ most recent generated frames up to and including the anchor frame $\rvx_t$. The interpolator then generates the $u$ frames of the current interval via

\begin{equation}p_\theta(\rvx_{t:t'} \mid \rvx_{t-p\Delta t : t}, \rvx_{t'}, \rva, t_{\text{rem}})= \prod_{k=1}^{u} p_\theta\bigl(\rvx_{t+k\Delta t} \;\big|\; \rvx_{t-p\Delta t : t+(k-1)\Delta t}, \rvx_{t'}, \rva, t_{\text{rem}}\bigr),\end{equation}

where $t_{\text{rem}} \equiv (u - k + 1)\Delta t$ is the temporal distance to the next anchor frame. For the very first interpolation interval (where no prior fine‑grained frames exist), we simply use a prefix consisting only of the initial anchor frame (i.e., $p=1$). This prefix mechanism ensures temporal continuity across anchor boundaries and eliminates discontinuities that would otherwise arise from restarting interpolation at each anchor frame independently.

\section{Experiments}\label{expriment}

In this section, we provide details on the application of \method to molecular dynamics (MD) trajectory generation.

\textbf{Dataset.}
We evaluate \method on two modalities: protein and protein-ligand. For protein dynamics, we adopt the mdCATH dataset~\cite{mirarchi2024mdcath}, which provides MD simulations for a wide range of protein domains. mdCATH features simulations of 5,398 domains at five different temperatures, each with five replicas. The production frames are sampled at 1ns intervals from simulations. We follow the dataset construction of TEMPO~\cite{xu2025tempo}, which focuses on the 320K trajectories and uses 1,000 domains for training, 50 for validation, and 64 for testing, with an average sequence similarity of 18.93\% between the training and test sets, ensuring a challenging generalization benchmark.

For protein-ligand dynamics, we use the MISATO dataset~\cite{siebenmorgen2024misato}, which provides trajectories capturing the coupled dynamics of proteins and ligands. The simulations are performed at 300 K for 8 ns, yielding 100 snapshots per trajectory. We preprocess MISATO following a procedure similar to Boltz‑2~\cite{passaro2025boltz}, resulting in 12,662 systems for training, 1,258 for validation, and 1,276 for testing (see Appendix~\ref{app:data_preprocess} for preprocessing details). Due to the computational cost of trajectory generation, we randomly sample 100 systems from the test set for final evaluation, aligning the test scale with Boltz‑2. The PDB IDs of the sampled test systems are provided in Appendix~\ref{app:data_preprocess}.

\textbf{Implementation details.}
We initialize the context embedding layer and the diffusion decoder with weights from Protenix~\cite{chen2025protenix}. During training on MD trajectories, the diffusion decoder is fine‑tuned end‑to‑end together with the newly introduced and randomly initialized geometric encoder $\gE_\theta$ and the state‑transition kernel $\gT_\theta$. Architectural details of these modules are provided in the Appendix~\ref{app:arch_details}. We adopt the dual‑scale generation strategy described in Section~\ref{subsec:trajectory_sampling}. On the mdCATH dataset, we train two separate models: one for coarse‑scale keyframe generation at a 20ns interval and another for fine‑scale interpolation at a 1ns interval. Similarly, for MISATO, we train a coarse model at 0.8ns and a fine model at 0.08ns. This yields four trained models in total; we leave the exploration of a unified model for future work. In the interpolation stage, the chunk stitching prefix length is set to $p=5$ for both datasets. Comprehensive training hyperparameter settings are listed in the Appendix~\ref{app:hyperparameters}.

\subsection{Protein Dynamics}

\begin{table*}[t]
    \centering
    \caption{\textbf{Evaluation on mdCATH}. The best scores are highlighted in \textbf{bold} and the second-best scores are \underline{underlined}. $r$: Pearson correlation; $J$: Jaccard similarity; $\mathcal{W}_2$: 2-Wasserstein distance. The MD (400ns) column reports oracle scores computed using one 400\,ns MD trajectory from the reference set for each test target.}
    \label{tab:mdCATH_results}
    \scalebox{0.8}{
    
    \begin{tabular}{l|ccccc>{\columncolor{gray!15}}c|c}
    \toprule[1.5pt]
    \textbf{Metrics} & \textbf{AlphaFlow} & \textbf{EBA} & \textbf{MDGen} & \textbf{TEMPO} & \textbf{ConfRover} & \textbf{\method} &  \textbf{MD (400ns)} \\ 
    \midrule[1.2pt]
    Pairwise RMSD $r$ $\uparrow$ & \underline{0.88} & 0.80 & 0.65 & 0.73 & 0.69 & \textbf{0.89} & 0.96 \\
    Global RMSF $r$ $\uparrow$ & \underline{0.86} & 0.74 & 0.67 & 0.61 & 0.62 & \textbf{0.87} & 0.95 \\
    Per-target RMSF $r$ $\uparrow$ & \underline{0.87} & 0.80 & 0.77 & 0.52 & 0.81 & \textbf{0.88} & 0.95 \\
    \midrule[1.2pt]
    Root mean $\mathcal{W}_2$-dist. $\downarrow$ & \textbf{3.06} & 4.04 & 3.86 & 4.98 & 4.52 & \underline{3.30} & 1.89 \\
    MD PCA $\mathcal{W}_2$-dist. $\downarrow$ & \underline{2.39} & 2.67 & 2.75 & 2.38 & 3.04 & \textbf{2.01} & 1.67 \\
    Joint PCA $\mathcal{W}_2$-dist. $\downarrow$ & \underline{2.82} & 3.42 & 3.54 & 4.71 & 4.15 & \textbf{2.79} & 1.75 \\
    \% PC-sim $>0.5$ $\uparrow$ & 30.7 & \underline{36.7} & 3.17 & 7.94 & 22.2 & \textbf{45.3} & 61.9 \\
    \midrule[1.2pt]
    Weak contacts $J$ $\uparrow$ & 0.56 & 0.40 & \underline{0.57} & 0.48 & 0.43 & \textbf{0.67} & 0.84 \\
    Transient contacts $J$ $\uparrow$ & \textbf{0.34} & 0.19 & \underline{0.22} & 0.20 & \underline{0.22} & \textbf{0.34} & 0.55 \\
    \midrule[1.2pt]
    t-RMSD Error $\downarrow$ & - & - & 2.71 & \underline{2.42} & 2.50 & \textbf{1.79} & 0.00 \\
    Decorr-TIC0 $r$ $\uparrow$ & - & - & \underline{0.07} & -0.14 & -0.05 & \textbf{0.25} & 1.00 \\
    MSM-Trans Error $\downarrow$ & - & - & \underline{0.28} & 0.31 & 0.30 & \textbf{0.24} & 0.23 \\
    \midrule[1.2pt]
    VAL-C$\alpha$ $\uparrow$ & 0.89 & \underline{0.91} & 0.07 & 0.01 & 0.53 & \textbf{0.97} & 0.94 \\
    \bottomrule[1.5pt]
    \end{tabular}

    }
    \vspace{-2mm}
\end{table*}

In Table~\ref{tab:mdCATH_results}, we compare \method against several recent trajectory generation approaches. For trajectory-level modeling, we evaluate MDGEN~\cite{jing2024mdgen}, TEMPO~\cite{xu2025tempo}, and ConfRover~\cite{shen2025confrover}. MDGEN and TEMPO are trained on the mdCATH dataset, while for ConfRover, which only provides inference code, we use the author-released checkpoint trained on the ATLAS dataset~\cite{vander2024atlas}. We exclude BioMD~\cite{feng2025biomd} and STAR-MD~\cite{shoghi2026scalable} from the comparison because neither codebase was publicly available at the time of writing. We additionally compare \method against ensemble generators, we include AlphaFlow~\cite{jing2024alphaflow} and EBA~\cite{lu2025eba}, both of which produce equilibrium ensembles without modeling temporal correlations. For a fair comparison, we use the AlphaFlow‑MD + Templates variant of AlphaFlow, which, like the trajectory methods, conditions on a single input structure.

For each protein in the test set, we condition on the initial structure and generate a trajectory of 400 frames at 1ns intervals, which we then compare to the corresponding MD reference trajectory. For AlphaFlow and EBA, we generate an independent set of conformations of equal size for comparison. 

Our evaluation framework spans three complementary aspects: ensemble-level properties, trajectory-specific kinetics, and structural validity. \textbf{For ensemble-level properties}, following AlphaFlow~\cite{jing2024alphaflow}, we assess three aspects: flexibility (Table~\ref{tab:mdCATH_results}, rows 2-4), distributional accuracy (Table~\ref{tab:mdCATH_results} rows 5-8), and ensemble observables (Table~\ref{tab:mdCATH_results} rows 9-10).
\textbf{For trajectory-specific kinetics}, we adopt three metrics: (i) t‑RMSD Error: For each frame, we compute the RMSD to the initial structure; the final metric is the root mean squared error between the predicted and reference RMSD sequences. It directly measures how well the model reproduces the magnitude of conformational change over time. (ii) Decorr-TIC0 $r$: The TIC0 decorrelation time is defined as the first lag at which the normalized autocorrelation of the first TICA component falls below 0.5. We report the Pearson correlation between predicted and reference decorrelation times. This tests whether the relaxation timescale is captured. (iii) MSM‑Transition Error: The metric is the root mean squared error between the transition matrices of the reference and predicted Markov state models (MSMs), defined on the same MSM state space. It quantifies how faithfully the model reproduces state‑to‑state kinetic probabilities.
\textbf{For structural validity}, we report VAL-C$\alpha$, the fraction of generated conformations in which no pair of C$\alpha$ atoms is either clashing or dissociated. 
Full details of all metrics are given in Appendix~\ref{app:eval_metrics}.

As shown in Table~\ref{tab:mdCATH_results}, \method consistently outperforms prior trajectory models across all evaluation dimensions, while matching or surpassing dedicated ensemble generators on most static distribution metrics. On ensemble‑level properties, \method reaches the highest Pearson correlations for both per‑target RMSF and aggregate flexibility measures (pairwise RMSD, global RMSF). The Joint PCA $\mathcal{W}_2$ distance and the fraction of principal component similarities above 0.5 are the best among all methods, indicating better reproduction of dominant collective motions~\cite{jing2024alphaflow}. Additionally, \method matches or surpasses AlphaFlow‑MD + Templates variant of AlphaFlow on both weak and transient contact Jaccard similarities. Overall, this demonstrates that \method generates conformational ensembles of comparable diversity and physical accuracy to dedicated ensemble generators.

The advantage of \method is most pronounced on trajectory‑specific kinetic metrics, which are inaccessible to static ensemble generators. \method reduces the t‑RMSD error to 1.79\AA, a significant improvement over the next best trajectory model. On the Decorr-TIC0 $r$, \method reaches $r=0.25$, markedly better than the weak or even negative correlations of prior methods, though still far from the strong correlation, highlighting the challenge of capturing slow kinetic modes. Similarly, the MSM transition error of 0.24 is the lowest among trajectory models and approaches the reference value of 0.23. Finally, \method achieves a structural validity score of 0.97, outperforming all baselines, indicating that the generated trajectories are essentially free of steric clashes or chain dissociation.

\begin{wraptable}{r}{0.49\linewidth}
\vspace{-2.0em}
    \caption{\textbf{Efficiency evaluation on the 1gnla01 protein (139 AA).} MD simulation is performed with Amber24 on GPU~\cite{salomon2013routine}, employing an integration time step of 4 fs.}
    \label{tab:efficiency_results}
    \begin{center}
    
    \begin{tabular}{l|cc}
    \toprule[1.5pt]
    & \makecell{Inference Time\\(Hour)} & \makecell{Peak Mem\\(GB)} \\ 
    \midrule[1.2pt]
    MD & $>$10 & $<$1 \\
    ConfRover & 0.467 & 22.53  \\
    \rowcolor{gray!15}
    \method & 0.174 & 3.445 \\
    \bottomrule[1.5pt]
    \end{tabular}

    \end{center}
    \vspace{-4mm}
\end{wraptable}

We compare trajectory-generation throughput on 1gnla01 for a
400\,ns trajectory with 400 output frames at 1\,ns intervals.
All methods are run on a single RTX 4090 GPU. As shown in
Table~\ref{tab:efficiency_results}, \method takes 0.174 hours,
compared with 0.467 hours for ConfRover and more than 10 hours
for MD, and uses about 15\% of ConfRover's peak GPU memory.
This comparison measures wall-clock time at a fixed trajectory
length and output frequency. The speedup at matched thermodynamic
or kinetic accuracy remains to be established.

\begin{figure*}[t]
    \centering
    \includegraphics[width=0.75\linewidth]{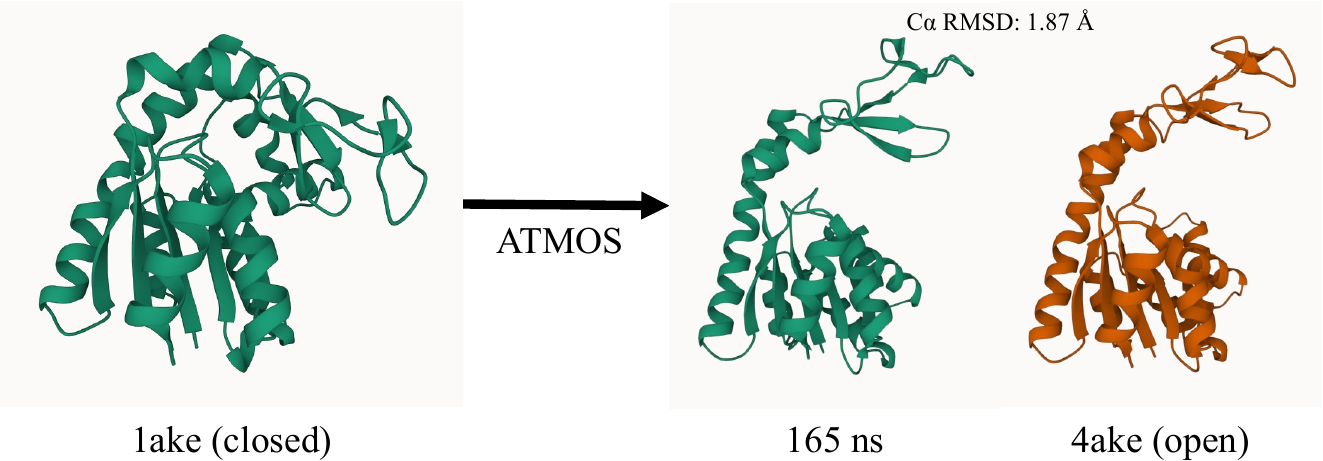}
    \caption{Case study on the adenylate kinase (AdK). \method simulates the domain motion of the adenylate kinase from close (0 ns) to open (165 ns). This protein is not included in the training set.}
    \label{fig:adk}
\end{figure*}

We further assess all-atom physical plausibility through
geometric validation and force-field energy evaluation on
16 randomly sampled mdCATH test systems. The evaluation
protocol and results are provided in
Appendix~\ref{app:physical_plausibility}
(Table~\ref{tab:physical_plausibility}). Additionally, we evaluate \method on Adenylate Kinase (AdK), a canonical system exhibiting large-scale, functionally relevant conformational changes between a closed (PDB: 1AKE) and an open (PDB: 4AKE) state. As shown in Figure~\ref{fig:adk}, starting from the closed state, trajectories generated by our model evolve toward the open state without explicit biasing forces, indicating that the model captures the intrinsic conformational landscape and coordinated domain motions underlying AdK function.

\subsection{Protein-Ligand Dynamics}

\begin{table*}[t]
    \centering
    \caption{\textbf{Evaluation on MISATO}. The best scores are highlighted in \textbf{bold} and the second-best scores are \underline{underlined}. $r$: Pearson correlation; $\mathcal{W}_2$: 2-Wasserstein distance.}
    \label{tab:misato_results}
    \scalebox{0.82}{
    
    \begin{tabular}{cl|cccc}
    \toprule[1.5pt]
    & \textbf{Method} & \textbf{per-target RMSF $r$} $\uparrow$ & \textbf{Root mean $\mathcal{W}_2$-dist} $\downarrow$ & \textbf{Steric Clashes} $\downarrow$ & \textbf{t-RMSD Error} $\downarrow$ \\
    \midrule[1.2pt]
    \multirow{3}{*}{Ligand} & NeuralMD & 0.574 & \underline{0.698} & 7.667 & 0.698 \\
    & DynamicBind & \underline{0.678} & 1.470 & 7.113 & - \\
    & Boltz-2 & 0.613 & 1.237 & \underline{0.210} & - \\
    \rowcolor{gray!15}
    \multicolumn{1}{c}{\cellcolor{white}{}}
    & \method & \textbf{0.769} & \textbf{0.693} & \textbf{0.000} & \textbf{0.522} \\
    \midrule[1.2pt]
    \multirow{2}{*}{\makecell{Protein-\\Ligand}} & DynamicBind & 0.193 & 2.288 & 52.74 & - \\
    & Boltz-2 & \underline{0.534} & \underline{1.992} & \underline{0.020} & - \\
    \rowcolor{gray!15}
    \multicolumn{1}{c}{\cellcolor{white}{}}
    & \method & \textbf{0.654} & \textbf{1.370} & \textbf{0.000} & \textbf{0.435} \\
    \bottomrule[1.5pt]
    \end{tabular}

    }
\end{table*}

For protein-ligand dynamics, we compare \method against NeuralMD~\cite{liu2025multi}, DynamicBind~\cite{lu2024dynamicbind}, and Boltz-2~\cite{passaro2025boltz}. NeuralMD models only ligand dynamics within a fixed protein environment, whereas DynamicBind and Boltz-2 generate protein-ligand conformational ensembles. Following the same protocol used for mdCATH, we condition each method on the initial complex structure and generate a complete trajectory. For ensemble generators, we sample an independent set of conformations of equal size. Given the short 8ns duration of MISATO trajectories and the localized nature of binding pocket motions, we evaluate four aspects: (1) flexibility prediction via per-target RMSF Pearson correlation, (2) distributional accuracy via root mean 2-Wasserstein distance, (3) structural quality via steric clashes, and (4) trajectory fidelity via t-RMSD error. Full metric definitions are provided in Appendix~\ref{app:eval_metrics}. We report separate results for ligand-only and whole-system evaluations where applicable. We note that Boltz-2 was trained on the complete MISATO dataset, which may constitute data leakage for our test set; we include it as a strong reference point.

As shown in Table~\ref{tab:misato_results}, \method achieves strong performance across both ligand-only and full protein-ligand evaluations. For ligand dynamics, \method attains the best scores in all four metrics. These results demonstrate that our model generates ligand motions closely matching the reference distribution and temporal evolution. On the full protein–ligand system, \method delivers the best scores in all four metrics. Although Boltz‑2 also shows competitive performance on some metrics, it is trained on the entire dataset (potential data leakage) and does not produce trajectories. In summary, \method not only excels at modeling ligand-in-pocket dynamics but also provides a strong framework for generating trajectories of the entire protein-ligand system.

\subsection{Ablation Studies}

\begin{table}[t]
    \centering
    \caption{\textbf{Contribution of pretrained Protenix components
    on mdCATH.} The best scores are highlighted in
    \textbf{bold} and the second-best scores are \underline{underlined}.}
    \label{tab:ablation_protenix}

    \small
    \setlength{\tabcolsep}{6pt}
    \begin{tabular}{l|cccc}
        \toprule[1.5pt]
        \textbf{Metrics}
        & \makecell{\textbf{ATMOS}\\(full)}
        & \makecell{\textbf{ATMOS}\\(ablated)}
        & \textbf{MDGen}
        & \textbf{ConfRover} \\
        \midrule[1.2pt]

        Pairwise RMSD $r$ $\uparrow$
        & \underline{0.89} & \textbf{0.90} & 0.65 & 0.69 \\

        Global RMSF $r$ $\uparrow$
        & \textbf{0.87} & \underline{0.86} & 0.67 & 0.62 \\

        \midrule
        Root mean $\mathcal{W}_2$-dist. $\downarrow$
        & \underline{3.30} & \textbf{2.99} & 3.86 & 4.52 \\

        \% PC-sim $>0.5$ $\uparrow$
        & \textbf{45.3} & \underline{40.3} & 3.17 & 22.2 \\

        \midrule
        Weak contacts $J$ $\uparrow$
        & \textbf{0.67} & 0.54 & \underline{0.57} & 0.43 \\

        Transient contacts $J$ $\uparrow$
        & \textbf{0.34} & \underline{0.30} & 0.22 & 0.22 \\

        \midrule
        t-RMSD Error $\downarrow$
        & \textbf{1.79} & \underline{2.01} & 2.71 & 2.50 \\

        Decorr-TIC0 $r$ $\uparrow$
        & \textbf{0.25} & \underline{0.134} & 0.07 & -0.05 \\

        MSM-Trans Error $\downarrow$
        & \textbf{0.24} & \underline{0.26} & 0.28 & 0.30 \\

        \midrule
        VAL-C$\alpha$ $\uparrow$
        & \textbf{0.97} & \underline{0.76} & 0.07 & 0.53 \\

        \bottomrule[1.5pt]
    \end{tabular}
\end{table}

\paragraph{Contribution of pretrained Protenix components.}
To assess the contribution of pretrained Protenix components,
we evaluate an ablated variant in which the 48-layer pretrained
Pairformer context trunk is removed and the diffusion decoder
is trained from scratch. These changes are applied to both
the coarse-scale keyframe generator and the fine-scale
interpolator.

As shown in Table~\ref{tab:ablation_protenix}, this variant
outperforms MDGen on nine of the ten reported metrics and
ConfRover on all ten. The full model further improves
structural validity (VAL-C$\alpha$: $0.76$ to $0.97$) and
kinetic accuracy (Decorr-TIC0 $r$: $0.134$ to $0.25$;
t-RMSD error: $2.01$ to $1.79$), whereas the ablated variant
achieves better pairwise RMSD correlation and root mean
$\mathcal{W}_2$ distance. These results suggest that the
gains of \method over these baselines cannot be attributed
solely to pretrained Protenix components.

\begin{table}[h]
    \centering
    \caption{\textbf{Ablation study on Chunk Stitching}.}
    \label{tab:ablation_chunk_stitching}
    \scalebox{0.8}{
    
    \begin{tabular}{c|ccccc}
    \toprule[1.5pt]
    $p$ & \textbf{Pairwise RMSD $r$} $\uparrow$ & \textbf{Root mean $\mathcal{W}_2$-dist.} $\downarrow$ & \textbf{t-RMSD Error} $\downarrow$ & \textbf{Decorr-TIC0 $r$} $\uparrow$ & \textbf{MSM-Trans Error} $\downarrow$ \\
    \midrule[1.2pt]
    0  & 0.87 & 2.55 & 1.83 & 0.10 & 0.25 \\
    5  & 0.89 & 3.30 & 1.79 & 0.25 & 0.24 \\
    10 & 0.90 & 2.99 & 1.80 & 0.20 & 0.24 \\
    \bottomrule[1.5pt]
    \end{tabular}
    
    }
\end{table}

\paragraph{Effect of chunk stitching.}
We examine the impact of the chunk stitching prefix length $p$ (defined in Section~\ref{subsec:trajectory_sampling}) on trajectory quality, comparing $p=0$ (i.e., restarting interpolation independently at each anchor frame), $p=5$, and $p=10$. As shown in Table~\ref{tab:ablation_chunk_stitching}, removing the prefix entirely ($p=0$) causes a sharp drop in the Decorr-TIC$_0$ correlation ($r$ falls from $0.25$ to $0.10$), confirming that temporal continuity across anchor boundaries is essential for faithfully capturing slow relaxation timescales. In contrast, the pairwise RMSD correlation and t‑RMSD error are largely robust to changes in $p$, indicating that overall flexibility and conformational drift are primarily determined by the keyframe generator.

Interestingly, $p=0$ yields a better Root mean $\mathcal{W}_2$ distance ($2.55$ vs $3.30$ for $p=5$). A possible explanation is that restarting interpolation at each anchor frame allows the interpolator to explore a broader conformational space between keyframes, inflating ensemble diversity at the cost of kinetic coherence. With $p=5$ and $p=10$, performance is nearly identical across all metrics, demonstrating that a short prefix of five fine‑grained frames already provides sufficient temporal context. 

\section{Conclusion}
We present \method, a unified generative framework for simulating biomolecular dynamics that generalizes from monomeric proteins to complex biomolecular systems. The model achieves state-of-the-art performance on large‑scale MD benchmarks, demonstrating its ability to generalize across diverse biological systems and paving the way toward foundational dynamics models. We provide further discussion of broader impacts and limitations in Appendix~\ref{app:impacts} and Appendix~\ref{app:limitations}, respectively.


\begin{ack}
We thank the anonymous reviewers for their valuable
comments and suggestions.

\paragraph{Funding Statement.}
This work was supported by BioGeometry through internship
funding, computational resources, and research guidance.
Liang Shi acknowledges financial aid from NeurIPS 2026
for conference participation.

\paragraph{Competing Interests Disclosures.}
The authors declare no competing interests.
\end{ack}

\bibliographystyle{abbrv}
\bibliography{reference}


\appendix

\section{Lay Summary}
Proteins are constantly moving, and their movements help determine how they work and interact with medicines. Conventional computer simulations can reveal this motion but are costly to run. We introduce ATMOS, an AI model that learns from existing simulations to generate molecular movies. Starting from a molecular structure, it keeps a running memory of earlier motion to guide what comes next. It first generates snapshots at widely spaced times, then fills in the motion between them. In tests on proteins and proteins interacting with small molecules, ATMOS reproduced several properties of simulated motion better than competing AI methods. It could help researchers explore molecular behavior more efficiently, although it does not yet replace conventional simulations or reliably capture very slow molecular changes.

\section{Broader Impacts} \label{app:impacts}
\method has significant broader impacts for computational biology and drug discovery. It provides a general‑purpose framework for modeling biomolecular dynamics that could accelerate therapeutic discovery and enhance biological understanding. However, ethical concerns, such as potential misuse in drug development, warrant careful consideration. Responsible deployment and transparency are essential to maximize benefits and mitigate risks.

\section{Limitations} \label{app:limitations}
While \method achieves strong performance across diverse benchmarks, several limitations point to promising directions for future work. First, our current approach trains separate models for different temporal scales with a fixed stride, rather than learning dynamics across a continuum of timescales within a single unified model. Second, for protein-ligand systems, the largest publicly available benchmark, MISATO, provides trajectories of only 8\,ns, which primarily probe local pocket flexibility rather than multi-state binding transitions or slow dissociation events. The absence of large, transferable, long-timescale protein-ligand MD datasets currently prevents a rigorous evaluation of such global kinetic phenomena. Third, we train and evaluate on each benchmark independently, as varying simulation protocols and temperature ranges across datasets hinder straightforward unification. Establishing a single dynamics foundation model that generalizes across systems, temperatures, and timescales remains an important objective that we leave for future investigation.

Our throughput benchmark fixes the trajectory length and output
frequency on a single protein. A fidelity-matched comparison with MD
is an important direction for future work. Such an evaluation would
measure the wall-clock cost of reaching comparable accuracy on
specified thermodynamic and kinetic observables.

\section{Architectural Details} \label{app:arch_details}
This section provides implementation details for the key modules introduced in Section~\ref{method:arch}.

\begin{figure*}[ht]
    \centering
    \includegraphics[width=1.0\linewidth]{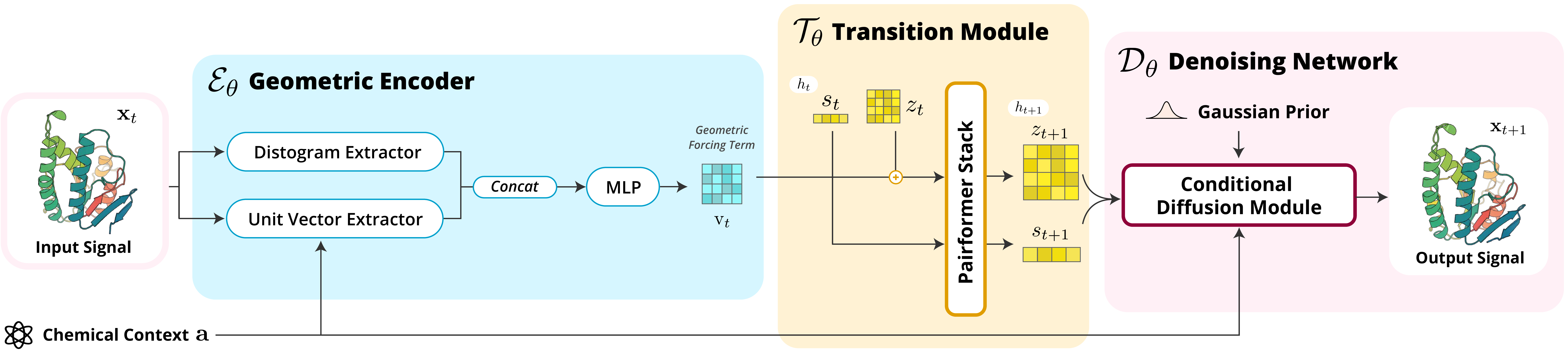}
    \caption{\textbf{Detailed architecture of our framework.}}
    \label{fig:flow}
\end{figure*}

\subsection{Context and Input Encoding}
We initialize the context embedding layer $\mathbf{e}_\theta$ with the pre‑trained trunk module of Protenix~\cite{chen2025protenix} (an open‑source reproduction of AlphaFold3~\cite{abramson2024af3}), which includes the Input Feature Embedder, MSA Module and Pairformer. This module is frozen during training, similar to ConfRover~\cite{shen2025confrover}.

The geometric encoder $\gE_{\theta}$, which is a MLP, maps the current conformation $X_t$ into a forcing term $\mathbf{v}_t$. $\mathbf{v}_t$ encodes pairwise geometric features.

\textbf{Pairwise geometric features.} For each token pair $(i,j)$, we compute invariant features that encode their relative geometry. For protein residues, the token center is taken as the C$\beta$ atom (C$\alpha$ for glycine); for ligand atoms, the atom itself serves as the center. We compute (1) a distance histogram between the two centers, and (2) unit vectors that express the displacement vector in the local coordinate frame of each token (constructed from backbone atoms for residues and neighboring atoms for ligands). These features are concatenated and projected:
\begin{equation}
\mathbf{v}_t[i,j] = \mathrm{MLP}\bigl(\mathrm{concat}(\text{Distogram}_{ij},\; \text{UnitVectors}_{ij})\bigr).
\end{equation}

\subsection{State Transition}\label{app:state_transition}
The core propagator of \method is the state transition function $\gT_\theta$, which evolves the latent state $\rvh$ from $t$ to $t+1$. We instantiate $\gT_\theta$ using the Pairformer module~\cite{abramson2024af3}. We use 4 Pairformer blocks in our implementation. The architectural hyperparameters are listed in Table~\ref{tab:pairformer_hyperparams}.

\begin{table}[h]
\centering
\caption{Architectural hyperparameters for the Pairformer.}
\label{tab:pairformer_hyperparams}
\begin{tabular}{lc}
\toprule
Hyperparameter & Value \\
\midrule
Number of Pairformer blocks & 4 \\
Single representation dimension & 384 \\
Pair representation dimension & 128 \\
Triangle multiplication hidden dimension & 128 \\
Number of triangle attention heads & 4 \\
Transition layer expanding factor & 4 \\
Pair attention dropout rate & 0.25 \\
\bottomrule
\end{tabular}
\end{table}

\subsection{Structure Decoding}\label{app:structure_decoding}
We model the generation of the next frame $\rvx_{t+1}$ as a reverse diffusion process, with a denoising network $\gD_{\theta}$ that predicts the clean structure from a noisy state. The diffusion decoder $\gD_\theta$ is initialized from the weights of Protenix~\cite{chen2025protenix}. Together with the context embedding layer $\mathbf{e}_\theta$ , this provides a prior on plausible biomolecular geometries, enabling data‑efficient learning of dynamics. We parameterize the decoder $\gD_\theta$ using an EDM‑style diffusion process~\cite{karras2022elucidating}. The conditional diffusion sampling procedure is detailed in Algorithm~\ref{alg:diffusion}. We use $N_{\text{step}}=50$ diffusion steps, a step scaling factor $\eta=1.5$, and a noise scaling schedule parameter $\gamma_0=0.8$ with a minimum $\gamma_{\text{min}}=1.0$. The noise scaling factor $\lambda$ is set to $1.5$ to encourage diversity.


\begin{algorithm}[t]
   \caption{Conditional Diffusion Sampling}
   \label{alg:diffusion}
\begin{algorithmic}[1]
    \STATE {\bfseries Input:} Diffusion steps $N_{\text{step}}$, latent state $\mathbf{h}$, chemical context $\mathbf{a}$, sampling hyperparameters $\lambda, \eta, \gamma_0, \gamma_{\text{min}}$.
    \STATE {\bfseries Output:} Generated coordinates $\mathbf{x}$.
    \STATE noise\_schedule $\gets$ inference\_noise\_scheduler($N_{\text{step}}$)
    \STATE $\mathbf{x}\gets\,$ noise\_schedule[0] $* \mathcal{N}(0, I)$ \# Initialize with noise
    \FOR {$i \gets 1\,\text{to}\,\text{len}(\text{noise\_schedule})\,$}
        \STATE $\gamma^{\text{last}}, \gamma \gets$ noise\_schedule[$i-1$], noise\_schedule[$i$]
        \STATE $\mathbf{x}\gets\,$ centre\_random\_augmentation($\mathbf{x}$) \# Apply random rigid augmentation
        \STATE $\gamma' \gets \gamma_0$ if $\gamma > \gamma_{\text{min}}$ else $0$
        \STATE $t_{\text{hat}} \gets \gamma^{\text{last}} * (\gamma' + 1)$
        \STATE $\mathbf{x}_{\text{noisy}} \gets \mathbf{x} + \lambda \sqrt{t_{\text{hat}}^2 - (\gamma^{\text{last}})^2} * \mathcal{N}(0, I)$
        \STATE $\mathbf{x}_{\text{denoised}} \gets$ $\gD_{\theta}$($\mathbf{x}_{\text{noisy}}, \mathbf{a}, \mathbf{h}, t_{\text{hat}}$) \# Conditional denoising
        \STATE $\delta \gets (\mathbf{x}_{\text{noisy}} - \mathbf{x}_{\text{denoised}}) / t_{\text{hat}}$
        \STATE $\Delta t \gets \gamma - t_{\text{hat}}$
        \STATE $\mathbf{x} \gets \mathbf{x}_{\text{noisy}} + \eta * \Delta t * \delta$
    \ENDFOR
    \STATE {\bfseries Return} $\mathbf{x}$
\end{algorithmic}
\end{algorithm}

\section{MISATO Dataset Preprocessing} \label{app:data_preprocess}
We preprocess the MISATO dataset following a procedure similar to Boltz‑2~\cite{passaro2025boltz}, resulting in 12,662 systems for training, 1,258 for validation, and 1,276 for testing.

The preprocessing consists of several steps. First, we download the MD restart file (to extract bond topology) and the MD result file (which contains trajectory coordinates and can be used to infer chain\_ids). Next, we generate a (.prmtop, .nc) pair for each protein–ligand complex. The .prmtop file holds topology information and the .nc file stores the trajectory coordinates. These files are loaded with MDTraj~\cite{mcgibbon2015mdtraj}, and each frame is exported as an individual PDB file. This method preserves complete topology (including bonds) and ensures a robust ligand representation, whereas directly constructing a PDB from the MD result file can lose atoms and introduce parsing inconsistencies.

We then add CCD (Chemical Component Dictionary) information to the loaded (.prmtop, .nc) pair, and convert them to (.pdb, .xtc) format. Systems that do not contain a MOL (ligand) residue are excluded. We also remove systems where the ligand’s atomic elements differ from those in the corresponding PDBBind MOL2 files, ensuring chemical consistency. Additionally, trajectories in which the ligand drifts more than 12\AA\ from the protein at any frame are filtered out. 

Furthermore, we remove approximately 500 systems that cannot be processed by the Protenix~\cite{chen2025protenix} data pipeline due to unexpected errors, indicating potential format, content, or geometry issues. (We first converts the PDB file to mmCIF format and then applies Protenix's standard processing.) The processed MISATO data and preprocessing code are publicly available; see Appendix~\ref{app:assets}.

The PDB IDs of the randomly sampled test systems are as follows: \texttt{16PK, 1AO8, 1D2S, 1DHJ, 1DRK, 1F73, 1IK4, 1JQD, 1NP0, 1WE2, 1XK9, 1Y2F, 1YYS, 2BET, 2EXG, 2GKL, 2HXM, 2I6B, 2LTO, 2NQI, 2PZE, 2QIC, 2RA6, 2RR4, 2V54, 2WJG, 2Y4K, 2Y4M, 2ZA3, 3BLU, 3C39, 3D9M, 3EOR, 3F69, 3G3N, 3GPO, 3HII, 3JUQ, 3MAG, 3ME9, 3MEU, 3QVU, 3S0B, 3TNE, 3ZOT, 4A22, 4ASK, 4AYR, 4B0C, 4DPU, 4DZY, 4G0Q, 4H7Q, 4I7B, 4KM2, 4KQO, 4KQR, 4KXL, 4L0L, 4MRE, 4NA4, 4NRQ, 4O1L, 4O24, 4ODL, 4UFH, 4X11, 4XAQ, 4YXD, 4ZB6, 5C13, 5C4K, 5ECT, 5EFC, 5F1R, 5H5Q, 5J6D, 5MXX, 5NPB, 5NPC, 5NPD, 5TPG, 5TT8, 5ZDC, 6BOE, 6C3U, 6CQ5, 6CW4, 6DYN, 6DYS, 6FAC, 6FAM, 6FNR, 6FNT, 6MIN, 6NP3, 6OA3, 6OIO, 6P12, 6PYA}.

\section{Training} \label{app:hyperparameters}
All experiments are conducted on 8 NVIDIA RTX 4090 GPUs with distributed data parallelism.

\textbf{Batching.} We employ a three‑level batching strategy. First, each GPU processes one independent trajectory per training step. Second, for each trajectory we sample four subsequences (with random start indices and fixed stride) and compute the loss over all four in parallel. Third, for each target frame in a subsequence we denoise 12 independent noise samples (diffusion batch size=12) to stabilize the diffusion objective.

\textbf{Optimization.} We use the Adam optimizer with a learning rate of $5\times10^{-4}$ and train for 40k steps.

\textbf{Cropping.} During training we crop each system to a fixed size of 256 tokens. For mdCATH we take a contiguous crop of 256 residues; for MISATO we apply a spatial interface crop that includes the ligand and its surrounding protein residues. 

\textbf{Loss weighting.} The training objective is a weighted sum of the reconstruction loss $\mathcal{L}_{\text{recon}}$ and the latent fidelity loss $\mathcal{L}_{\text{disto}}$ (defined in Section~\ref{method:training_objectives}). We set $\lambda_{\text{recon}} = 4.0$ and $\lambda_{\text{disto}} = 0.03$. The reconstruction loss also includes an atom‑type weight $w_{\text{entity}}(i)$, set to $11.0$ for ligand atoms and $1.0$ for protein atoms. We use $\sigma_{\text{data}} = 16$ following AlphaFold 3~\cite{abramson2024af3}.

\section{Evaluation Metrics} \label{app:eval_metrics}

\subsection{mdCATH} \label{app:eval_metrics:mdCATH}

\subsubsection{Ensemble-level Properties}

Following AlphaFlow~\cite{jing2024alphaflow}, we evaluate the generated ensembles from three aspects: predicting flexibility, distributional accuracy, and ensemble observables. For each protein, the generated trajectory is compared against all replicas of the corresponding MD simulations.

\textbf{Predicting flexibility.} We assess the ability of a model to reproduce the dynamic fluctuations observed in MD simulations using three correlation-based metrics:

\begin{itemize}
    \item \textbf{Pairwise RMSD $r$}: For each protein, we compute the average C$\alpha$-RMSD between every pair of conformations in the ensemble, which serves as a scalar measure of overall flexibility. The Pearson correlation between these pairwise RMSD values and those of the reference MD ensemble is reported.
    
    \item \textbf{Global RMSF $r$}: We calculate the root-mean-square fluctuation (RMSF) of each residue (represented by its C$\alpha$ atom) across the ensemble. All per-residue RMSF values from all test proteins are concatenated, and the Pearson correlation with the reference RMSF values is computed.
    
    \item \textbf{Per-target RMSF $r$}: For each individual protein, we compute the Pearson correlation between the predicted and reference per-residue RMSF vectors. The median of these per-protein correlations across the test set is reported.
\end{itemize}

\textbf{Distributional accuracy.} We assess the model's ability to reproduce the correct distribution of atomic positions using four metrics based on the 2-Wasserstein distance ($\mathcal{W}_2$):

\begin{itemize}
    \item \textbf{Root mean $\mathcal{W}_2$-dist}: The root mean Wasserstein distance (RMWD) between ensembles is defined as:
    \begin{equation}
        \text{RMWD}(\mathcal{X}, \mathcal{Y}) = \sqrt{\frac{1}{N}\sum^{N}_{i = 1}\mathcal{W}^2_2(\mathcal{N}[\mathcal{X}_i], \mathcal{N}[\mathcal{Y}_i])},
    \end{equation}
    where $\mathcal{N}[\mathcal{X}_i]$ denotes a 3D Gaussian fitted to the positional distribution of the $i$-th atom in ensemble $\mathcal{X}$. We report the median value over the test set.

    \item \textbf{MD PCA $\mathcal{W}_2$-dist}: To assess the similarity of collective motions, we project the C$\alpha$ atom positions of the generated ensemble onto the first two principal components (PCs) derived from the reference MD ensemble. The 2-Wasserstein distance (in Angstrom units of RMSD) between the projected distributions of the generated and reference ensembles in this 2D PCA space is computed and reported.

    \item \textbf{Joint PCA $\mathcal{W}_2$-dist}: We perform PCA on the combined set of C$\alpha$ atom positions from the generated and reference conformations (equally weighted) to obtain a joint set of principal components. Both ensembles are then projected onto the first two joint PCs, and the 2-Wasserstein distance between their projected C$\alpha$ distributions is computed. 

    \item \textbf{\% PC-sim $>$ 0.5}: For each protein, we compute the unsigned cosine similarity between the top principal component (derived from C$\alpha$ positions) of the generated ensemble and that of the reference MD ensemble. A value exceeding 0.5 indicates that the dominant collective motion is successfully captured. We report the percentage of test proteins for which this condition holds.
\end{itemize}

\textbf{Ensemble observables.} 
To evaluate whether the model captures biologically relevant contact dynamics, we compute two sets of inter‑residue contacts and compare them to the reference MD ensemble:

\begin{itemize}
    \item \textbf{Weak contacts $J$:} 
    We identify pairs of C$\alpha$ atoms that are in contact (distance $\leq 8$\AA) in the crystal structure but dissociate (distance $> 8$\AA) in more than 10\% of the ensemble structures. The Jaccard similarity between the set of such weak contacts in the generated ensemble and that in the reference MD ensemble is reported.

    \item \textbf{Transient contacts $J$:} 
    Conversely, we identify pairs of C$\alpha$ atoms that are not in contact (distance $> 8$\AA) in the crystal structure but associate (distance $\leq 8$\AA) in more than 10\% of the ensemble structures. The Jaccard similarity between the generated and reference sets of transient contacts is reported.
\end{itemize}

\subsubsection{Trajectory-specific Kinetics}

\textbf{t-RMSD Error.}
We report the t-RMSD Error, a trajectory‑level metric that quantifies a model’s ability to reproduce the magnitude of conformational change over time.  This metric is computed by (1) calculating the RMSD of each frame to the initial frame, and (2) computing the RMSE between the resulting RMSD sequences of the predicted and ground‑truth trajectories.

\textbf{Decorr-TIC0 $r$.}
To apply TICA, each frame of a trajectory is featurized by the sine and cosine of the backbone $\phi$ and $\psi$ dihedral angles. These features are then projected onto the first two time-lagged independent components (TIC0 and TIC1) using the \texttt{deeptime}~\cite{hoffmann2021deeptime} library, with a lag time of 1ns.For each trajectory, we compute the normalized autocorrelation function of the first component (TIC0) via \texttt{statsmodels}~\cite{seabold2010statsmodels}. The TIC0 decorrelation time is defined as the smallest lag (in frames) at which this autocorrelation drops below~0.5. Finally, we report the Pearson correlation coefficient between the decorrelation times of the predicted and reference trajectories across the test set, denoted as Decorr-TIC0 $r$. A higher correlation indicates better agreement in the timescale of the dominant slow relaxation process.

\textbf{MSM-Transition Error.}
We first cluster the two-dimensional features (TIC0 and TIC1) of the reference trajectory into 100 microstates using the KMeans++~\cite{arthur2007k} algorithm.
A Markov state model (MSM) is then constructed from the microstate assignments with a lag time of 1~ns, and the PCCA+~\cite{roblitz2013fuzzy} algorithm coarse‑grains these microstates into 10 metastable states.
To ensure full connectivity, any microstates classified as inactive are reassigned to the nearest active metastable cluster.
The resulting clustering and PCCA+ models, derived exclusively from the reference trajectory, define a fixed mapping from the TIC space into the 10 metastable states.

We apply this mapping to both the generated and reference trajectories.
For each trajectory, transition counts between metastable states are tallied over consecutive frames at a 1~ns lag, producing a transition count matrix, which is then normalized to a row‑stochastic transition probability matrix, denoted $\mathbf{T}_{\text{pred}}$ and $\mathbf{T}_{\text{ref}}$, respectively.
The MSM‑Transition Error is the root mean squared error between these two matrices. A lower error indicates that the generated dynamics reproduce the reference state‑to‑state transition probabilities more faithfully.

\subsubsection{Structural Validity}

\textbf{VAL-C$\alpha$.}
We assess the structural stability of generated protein conformations over long timescales using a geometric validity criterion. Following ConfDiff~\cite{wang2024confdiff}, a conformation is deemed valid if two conditions hold: (I) no pair of C$\alpha$ atoms is closer than $3.0\,\mathrm{\AA}$ (steric clash check), and (II) the C$\alpha$–C$\alpha$ distance of each pair of consecutive residues does not exceed $4.19\,\mathrm{\AA}$ (backbone continuity check). We report the fraction of frames in a trajectory that satisfy both conditions, denoted VAL‑C$\alpha$.

\subsection{MISATO}
The MISATO dataset captures the coupled dynamics of protein–ligand complexes, with the ligand's motion confined to the protein's binding pocket. For evaluation, we crop the pocket–ligand region. The pocket is defined as all residues where any heavy atom lies within 10\AA\ of any ligand heavy atom, following the threshold used in AlphaFold 3~\cite{abramson2024af3}.

We assess four aspects: flexibility prediction (per‑target RMSF Pearson correlation), distributional accuracy (root mean 2‑Wasserstein distance), steric clashes, and trajectory accuracy (t-RMSD Error). Except for steric clashes, the definitions align with those in Section~\ref{app:eval_metrics:mdCATH}, but extended to include all atoms (protein and ligand).

Steric clashes are reported as the average number of clashes per generated trajectory. For ligand‑only evaluation, we count clashes between ligand heavy atoms. For whole‑complex evaluation, we count clashes between protein heavy atoms and ligand heavy atoms. A clash is defined as two heavy atoms being closer than 1.1\AA\ again following AlphaFold 3~\cite{abramson2024af3}.

\section{Additional Physical Plausibility Evaluation}
\label{app:physical_plausibility}

\paragraph{Evaluation protocol.}
To complement the structural validity evaluation in the main
paper, we assess the all-atom physical plausibility of generated
structures using direct geometric and energetic checks.
Because these checks are computationally expensive when applied
frame by frame, we evaluate them on 16 randomly sampled mdCATH
test systems. All reported values are means.

We use MolProbity~\cite{williams2018molprobity} to evaluate
Ramachandran outliers and favored regions, rotamer outliers,
root-mean-square deviations of bond lengths and bond angles
from ideal geometry, and clashscore, which measures severe
steric overlaps per 1,000 atoms. In addition, we use
OpenMM~\cite{eastman2024openmm} to compute force-field energies.
For energy evaluation, hydrogens and OXT atoms are added,
and a short energy minimization with
\texttt{maxIterations=100} is performed before computing
the energies. The same protocol is applied to all methods
and the MD reference.

\begin{table}[t]
    \centering
    \caption{\textbf{Additional physical plausibility metrics
    on mdCATH.} Evaluation is performed on 16 randomly sampled
    test systems, and all values are reported as means.
    The best scores among the evaluated generative methods,
    excluding the MD reference, are highlighted in
    \textbf{bold}. Rama.\ denotes Ramachandran.
    Force-field energies are evaluated after a short energy
    minimization with \texttt{maxIterations=100}.}
    \label{tab:physical_plausibility}

    \resizebox{\linewidth}{!}{
    \begin{tabular}{l|ccccccc}
        \toprule[1.5pt]
        \textbf{Method}
        & \makecell{Rama.\\outliers (\%) $\downarrow$}
        & \makecell{Rama.\\favored (\%) $\uparrow$}
        & \makecell{Rotamer\\outliers (\%) $\downarrow$}
        & \makecell{RMS\\(bonds) $\downarrow$}
        & \makecell{RMS\\(angles) $\downarrow$}
        & \makecell{Clashscore\\$\downarrow$}
        & \makecell{Force-field\\energy $\downarrow$} \\
        \midrule[1.2pt]

        MD reference
        & 1.100 & 92.36 & 2.863 & 0.036 & 3.654
        & 0.953 & $-7962$ \\

        \midrule[1.2pt]

        AlphaFlow
        & 0.617 & 96.72 & 0.776 & 0.031 & 2.350
        & 25.13 & $-7464$ \\

        EBA
        & 0.202 & \textbf{97.61} & 1.328 & 0.100 & 4.394
        & 9.413 & $1.5 \times 10^{8}$ \\

        MDGen
        & 7.944 & 78.70 & 4.799 & 0.085 & 6.112
        & 247.7 & $2 \times 10^{21}$ \\

        ConfRover
        & 3.955 & 89.48 & 4.143 & 0.053 & 4.165
        & 147.3 & $3 \times 10^{26}$ \\

        TEMPO
        & 11.66 & 20.05 & N/A & 0.372 & 22.61
        & 476.1 & N/A \\

        \rowcolor{gray!15}
        \textbf{ATMOS}
        & \textbf{0.192}
        & 97.48
        & \textbf{0.394}
        & \textbf{0.014}
        & \textbf{1.786}
        & \textbf{6.204}
        & $\mathbf{-7931}$ \\

        \bottomrule[1.5pt]
    \end{tabular}
    }
\end{table}

\paragraph{Results and interpretation.}
As shown in Table~\ref{tab:physical_plausibility},
\method achieves the best scores among the evaluated
generative methods on most geometric metrics, including
Ramachandran outliers, rotamer outliers, bond-length and
bond-angle deviations, and clashscore. Its mean force-field
energy after minimization is close to that of the MD
reference ($-7931$ versus $-7962$), although its clashscore
remains higher ($6.204$ versus $0.953$).

These results provide complementary evidence that \method
improves all-atom physical plausibility in terms of local
geometry, steric consistency, and force-field energy under
the same evaluation protocol. These checks do not, by
themselves, establish full thermodynamic correctness,
accurate long-timescale sampling, or the ability to replace
explicit MD simulations.

\section{Code, Data, and Model Availability}
\label{app:assets}

\paragraph{Released assets.}
The ATMOS source code and instructions for installation, training,
inference, and evaluation are available at
\url{https://github.com/VectorShi/ATMOS}.
Four trained model checkpoints and their configurations are available at
\url{https://modelscope.cn/models/SmartVector/ATMOS}.
Processed mdCATH and MISATO trajectories, loader indices, data splits,
and precomputed protein MSAs are available at
\url{https://modelscope.cn/datasets/SmartVector/ATMOS-dataset}.
The repository documents the asset layout, preprocessing procedures,
and evaluation conventions in \texttt{docs/assets.md} and the
dataset-specific guides.

\paragraph{Licenses and attribution.}
ATMOS builds on Protenix~\cite{chen2025protenix} and uses its
v0.5.0 initialization weights. The Protenix code and these model
parameters are distributed under Apache-2.0.
The source mdCATH~\cite{mirarchi2024mdcath} and
MISATO~\cite{siebenmorgen2024misato} datasets are distributed
under CC BY 4.0; our processed subsets retain attribution
to their original creators.
ATMOS source code is distributed under Apache-2.0, except
for components covered by their own licenses. In particular,
the trajectory conversion script adapted from MISATO is
distributed under LGPL-2.1-or-later.
Original copyright and license notices are retained in the
repository and summarized in \texttt{THIRD\_PARTY\_NOTICES.md}.

PDBbind v2020~\cite{wang2005pdbbind} is used during MISATO preprocessing and must
be obtained separately under the data provider's terms.
The release does not redistribute the original PDBbind files.
Source-code licensing does not replace the applicable terms
for separately distributed datasets, MSAs, model weights,
or common chemical data.


\newpage
\section*{NeurIPS Paper Checklist}

\begin{enumerate}

\item {\bf Claims}
    \item[] Question: Do the main claims made in the abstract and introduction accurately reflect the paper's contributions and scope?
    \item[] Answer: \answerYes{} 
    \item[] Justification: Yes. Our main contributions are detailed in Sec.~\ref{intro}. Also see Sec.~\ref{expriment} for more experimental evidence.

\item {\bf Limitations}
    \item[] Question: Does the paper discuss the limitations of the work performed by the authors?
    \item[] Answer: \answerYes{} 
    \item[] Justification: Yes, please see Appendix.~\ref{app:limitations} for limitations.

\item {\bf Theory assumptions and proofs}
    \item[] Question: For each theoretical result, does the paper provide the full set of assumptions and a complete (and correct) proof?
    \item[] Answer: \answerNA{} 
    \item[] Justification: The paper does not include theoretical results.

    \item {\bf Experimental result reproducibility}
    \item[] Question: Does the paper fully disclose all the information needed to reproduce the main experimental results of the paper to the extent that it affects the main claims and/or conclusions of the paper (regardless of whether the code and data are provided or not)?
    \item[] Answer: \answerYes{} 
    \item[] Justification: Yes. We provide a clear and detailed description of the architecture in Sec.~\ref{method} and Appendix~\ref{app:arch_details}, along with the necessary training configurations in Sec.~\ref{app:hyperparameters}. All datasets used in our experiments are also publicly accessible.

\item {\bf Open access to data and code}
    \item[] Question: Does the paper provide open access to the data and code, with sufficient instructions to faithfully reproduce the main experimental results, as described in supplemental material?
    \item[] Answer: \answerYes{} 
    \item[] Justification: The ATMOS code and execution instructions
are publicly available on GitHub, and the trained checkpoints
and processed datasets are available on ModelScope.
Appendix~\ref{app:assets} provides the corresponding links.
The supplied scripts cover ATMOS training, inference, and
benchmark evaluation on mdCATH and MISATO.

\item {\bf Experimental setting/details}
    \item[] Question: Does the paper specify all the training and test details (e.g., data splits, hyperparameters, how they were chosen, type of optimizer) necessary to understand the results?
    \item[] Answer: \answerYes{} 
    \item[] Justification: Please see Sec.~\ref{expriment} and Sec.~\ref{app:hyperparameters}.

\item {\bf Experiment statistical significance}
    \item[] Question: Does the paper report error bars suitably and correctly defined or other appropriate information about the statistical significance of the experiments?
    \item[] Answer: \answerNo{} 
    \item[] Justification: Following previous work, we do not report error bars.

\item {\bf Experiments compute resources}
    \item[] Question: For each experiment, does the paper provide sufficient information on the computer resources (type of compute workers, memory, time of execution) needed to reproduce the experiments?
    \item[] Answer: \answerYes{} 
    \item[] Justification: Yes. Please see Sec.~\ref{app:hyperparameters}.
    
\item {\bf Code of ethics}
    \item[] Question: Does the research conducted in the paper conform, in every respect, with the NeurIPS Code of Ethics \url{https://neurips.cc/public/EthicsGuidelines}?
    \item[] Answer: \answerYes{} 
    \item[] Justification: We followed the NeurIPS Code of Ethics.

\item {\bf Broader impacts}
    \item[] Question: Does the paper discuss both potential positive societal impacts and negative societal impacts of the work performed?
    \item[] Answer: \answerYes{} 
    \item[] Justification: Yes. Please see Appendix.~\ref{app:impacts}
    
\item {\bf Safeguards}
    \item[] Question: Does the paper describe safeguards that have been put in place for responsible release of data or models that have a high risk for misuse (e.g., pre-trained language models, image generators, or scraped datasets)?
    \item[] Answer: \answerNA{} 
    \item[] Justification: We do not foresee any high risk for misuse of this work.

\item {\bf Licenses for existing assets}
    \item[] Question: Are the creators or original owners of assets (e.g., code, data, models), used in the paper, properly credited and are the license and terms of use explicitly mentioned and properly respected?
    \item[] Answer: \answerYes{} 
    \item[] Justification:  We cite the original creators and specify
the licenses and access conditions of the principal assets in
Appendix~\ref{app:assets}. The code repository retains upstream
copyright and license notices and documents third-party
components.


\item {\bf New assets}
    \item[] Question: Are new assets introduced in the paper well documented and is the documentation provided alongside the assets?
    \item[] Answer: \answerYes{} 
    \item[] Justification: We release the ATMOS source code,
four trained model checkpoints with configurations, and
processed mdCATH and MISATO data with MSA features and split
indices. The accompanying documentation describes asset
organization, preprocessing, training, inference, and evaluation.
See Appendix~\ref{app:assets} for access and attribution,
and Appendix~\ref{app:limitations} for limitations.

\item {\bf Crowdsourcing and research with human subjects}
    \item[] Question: For crowdsourcing experiments and research with human subjects, does the paper include the full text of instructions given to participants and screenshots, if applicable, as well as details about compensation (if any)? 
    \item[] Answer: \answerNA{} 
    \item[] Justification: The paper does not involve crowdsourcing nor research with human subjects.

\item {\bf Institutional review board (IRB) approvals or equivalent for research with human subjects}
    \item[] Question: Does the paper describe potential risks incurred by study participants, whether such risks were disclosed to the subjects, and whether Institutional Review Board (IRB) approvals (or an equivalent approval/review based on the requirements of your country or institution) were obtained?
    \item[] Answer: \answerNA{} 
    \item[] Justification: The paper does not involve crowdsourcing nor research with human subjects.

\item {\bf Declaration of LLM usage}
    \item[] Question: Does the paper describe the usage of LLMs if it is an important, original, or non-standard component of the core methods in this research? Note that if the LLM is used only for writing, editing, or formatting purposes and does \emph{not} impact the core methodology, scientific rigor, or originality of the research, declaration is not required.
    \item[] Answer: \answerNA{} 
    \item[] Justification: The core method development in this research does not involve LLMs as any important, original, or non-standard components.

\end{enumerate}

\end{document}